\documentclass[%
aip,
rsi
 amsmath,amssymb,
 reprint,%
]{revtex4-1}

\usepackage{graphicx}
\usepackage{dcolumn}
\usepackage{bm}
\usepackage{xcolor}
\usepackage[utf8]{inputenc}
\usepackage[T1]{fontenc}
\usepackage{mathptmx}
\usepackage[switch, modulo]{lineno}
\usepackage{tikz}
\usepackage{tabularx,booktabs}
\newcolumntype{Y}{>{\centering\arraybackslash}X} 
\usepackage{array}

\begin{document}

\preprint{AIP/123-QED}

\title{
 Broad frequency tuning of a Nb$_{3}$Sn superconducting microwave cavity for dark matter searches \\
}


\author{D.~Maiello}\email{dora.maiello@phd.unipd.it} \affiliation{Dipartimento di Fisica e Astronomia, 35131 Padova, Italy}\affiliation{INFN, Sezione di Padova, 35131 Padova, Italy}
\author{R.~Di Vora} \affiliation{INFN, Laboratori Nazionali di Legnaro, 35020 Legnaro, Padova, Italy}
\author{D.~Ahn} \affiliation{INFN, Sezione di Padova, 35131 Padova, Italy}
\author{G.~Carugno} \affiliation{INFN, Sezione di Padova, 35131 Padova, Italy}
\author{R.~Cervantes} \affiliation{Superconducting Quantum Materials and Systems center (SQMS), Fermi National Accelerator Laboratory, Batavia, IL, 60510, USA}
\author{B.~Giaccone} \affiliation{Superconducting Quantum Materials and Systems center (SQMS), Fermi National Accelerator Laboratory, Batavia, IL, 60510, USA}
\author{A.~Ortolan} \affiliation{INFN, Laboratori Nazionali di Legnaro, 35020 Legnaro, Padova, Italy}
\author{S.~Posen} \affiliation{Fermi National Accelerator Laboratory, Batavia, IL, 60510, USA}
\author{G.~Ruoso} \affiliation{INFN, Laboratori Nazionali di Legnaro, 35020 Legnaro, Padova, Italy}
\author{G.~Sardo Infirri} \affiliation{Dipartimento di Fisica e Astronomia, 35131 Padova, Italy}\affiliation{INFN, Sezione di Padova, 35131 Padova, Italy}
\author{B.~Tennis} \affiliation{Fermi National Accelerator Laboratory, Batavia, IL, 60510, USA}
\author{S.~Tocci} \affiliation{INFN, Laboratori Nazionali di Frascati, 00044 Frascati, Roma, Italy}
\author{C.~Braggio} \email{caterina.braggio@unipd.it}\affiliation{Dipartimento di Fisica e Astronomia, 35131 Padova, Italy}\affiliation{INFN, Sezione di Padova, 35131 Padova, Italy} 

\AddToHook{shipout/firstpage}{%
  \begin{tikzpicture}[remember picture,overlay]
    \node[anchor=north east, inner sep=10pt] at (current page.north east)
      {\normalsize\bfseries FERMILAB-PUB-26-0145-SQMS-TD};
  \end{tikzpicture}
}

\date{\today}

\begin{abstract}
We demonstrate a novel broad-frequency tuning mechanism for superconducting microwave cavities designed for dark matter searches. Using a Nb$_3$Sn-coated cigar-shaped cavity operating at approximately 9\,GHz, we achieve continuous frequency tuning exceeding 1\,GHz by mechanically separating the two cavity halves: a "tuning-by-opening" technique. 
Finite-element method simulations predict that radiative losses do not degrade the quality factor even for large openings, as a closed cavity with an intrinsic quality factor of $10^7$ maintains this value for apertures up to 9\,mm, corresponding to a tuning range from 9.0 to 7.5\,GHz. Experimental validation using both copper ring spacers and a continuous sliding mechanism confirms $Q_0$ values exceeding the dark matter quality factor across the entire explored frequency range, despite mechanical imperfections and film non-uniformities on the lateral plates. This tuning approach avoids inserting elements into the resonant volume, making it particularly suitable for high-Q superconducting cavities in axion haloscope experiments and readily applicable to REBCO-based implementations capable of operating in multi-tesla magnetic fields.
\end{abstract}

\maketitle

\section{\label{sec:intro} Introduction}
The nature of dark matter (DM) is, to this day, still elusive\cite{Bertone2018}.
A widely accepted hypothesis is that DM consists of massive particles beyond the Standard Model, moving at non-relativistic velocities, with a local Galactic density \cite{Read2014} of $\rho_{\mathrm{DM}} \approx 0.45\,\mathrm{GeV/cm^3}$.

Well-motivated candidates are axions and axion-like particles (ALPs), which can convert to real photons in the presence of a magnetic field \cite{Sikivie1983}.
Another viable DM candidate is the dark photon, which interacts with the Standard Model photon through kinetic mixing characterized by a dimensionless parameter $\chi$. Since dark photons couple directly to electromagnetic fields no external magnetic field is required for their conversion \cite{Caputo:2021}.

As the expected signal power is extremely suppressed due to the small model-dependent coupling, haloscopes exploit the resonant enhancement within high quality factor microwave cavities \cite{Kim:2020, Cervantes:2024}.
There are significant worldwide experimental efforts to probe this interaction, that are mainly based on copper cavities with typical intrinsic quality factors $Q \sim 10^{4}$--$10^{5}$, including ADMX~\cite{ADMX:2025}, HAYSTAC~\cite{HAYSTAC:2025}, and CAPP's main axion experiment~\cite{CAPP:2024}. Better sensitivity can be achieved with larger quality factors enabled by superconducting cavities.
An existing accelerator superconducting radio frequency (SRF) cavity with $Q_0 \approx 10^{11}$ was used to obtain the deepest exclusion limit to dark photons~\cite{Cervantes:2024} at $f = 1.3\,$GHz. This is a TESLA-shaped~\cite{TESLA} single-cell niobium cavity, that cannot be operated with magnetic fields exceeding the niobium upper critical field $H_{\mathrm{c2}}$ of about $0.2 - 0.5\,$T \cite{Prozorov}, well below the multi-tesla field required in axion haloscope experiments. Dedicated studies have investigated superconducting materials with high $H_\mathrm{c2}$ like Nb$_3$Sn~\cite{Posen:2023} and NbTi~\cite{Alesini:2019} films and implementations based on REBCO (Rare-earth barium copper oxide)~\cite{Ahn:2022, Golm:2022}.

Wide-band searches require tunability: the mass range accessible to haloscope experiments spans from about $1\,\mu$eV to $100\,\mu$eV, requiring the use of different resonant cavities covering complementary frequency bands, and within each band the cavity resonance must be tuned. In copper cavities, the most widely adopted tuning approach relies on movable metallic or dielectric rods inserted into the cavity volume. However, rod‑based tuning is more challenging to implement in superconducting cavities without degrading the high quality factor, especially at high frequency, either due to radiative losses or to the low surface quality \cite{Otten:2016} of the rod, when using REBCO tapes.
Motivated by these limitations, alternative concepts have recently begun to emerge, including clamshell-type mechanisms demonstrated within the QUAX experiment~\cite{Braggio:2023,DiVora:2025,Infirri:2025} and cavity-opening strategies explored by the RADES collaboration~\cite{Golm2024}, which avoid inserting tuning elements into the resonant volume and are therefore particularly attractive for superconducting resonators.

Despite the progress in superconducting cavity performance, the active tuning range remains rather limited. 
For instance, by varying the pressure of the liquid helium bath, a tuning range of about 300\,kHz at 8.84\,GHz was obtained during a REBCO-cavity test with $Q_0\approx 6.5\times10^4$ under 11.7\,T magnetic field ~\cite{Ahyoune2025}. 
In a hybrid NbTi copper cavity for axion searches at 7.4\,GHz \cite{Marconato:2024,Braggio:2025}, 10\,MHz frequency tuning range can be accomplished by insertion of a triplet of sapphire rods without impacting the $10^6$ quality factor measured under 2\,T magnetic field. 
The SHANHE collaboration performed a dark photon search using a Nb SRF cavity with $Q \approx10^{11}$, achieving a tuning range of $1.37\,\mathrm{MHz}$ around $1.3\,\mathrm{GHz}$ through mechanical compression of the cavity structure \cite{SHANHE:2024}.

\begin{figure}[t]
    \centering
    \includegraphics[width=\linewidth]{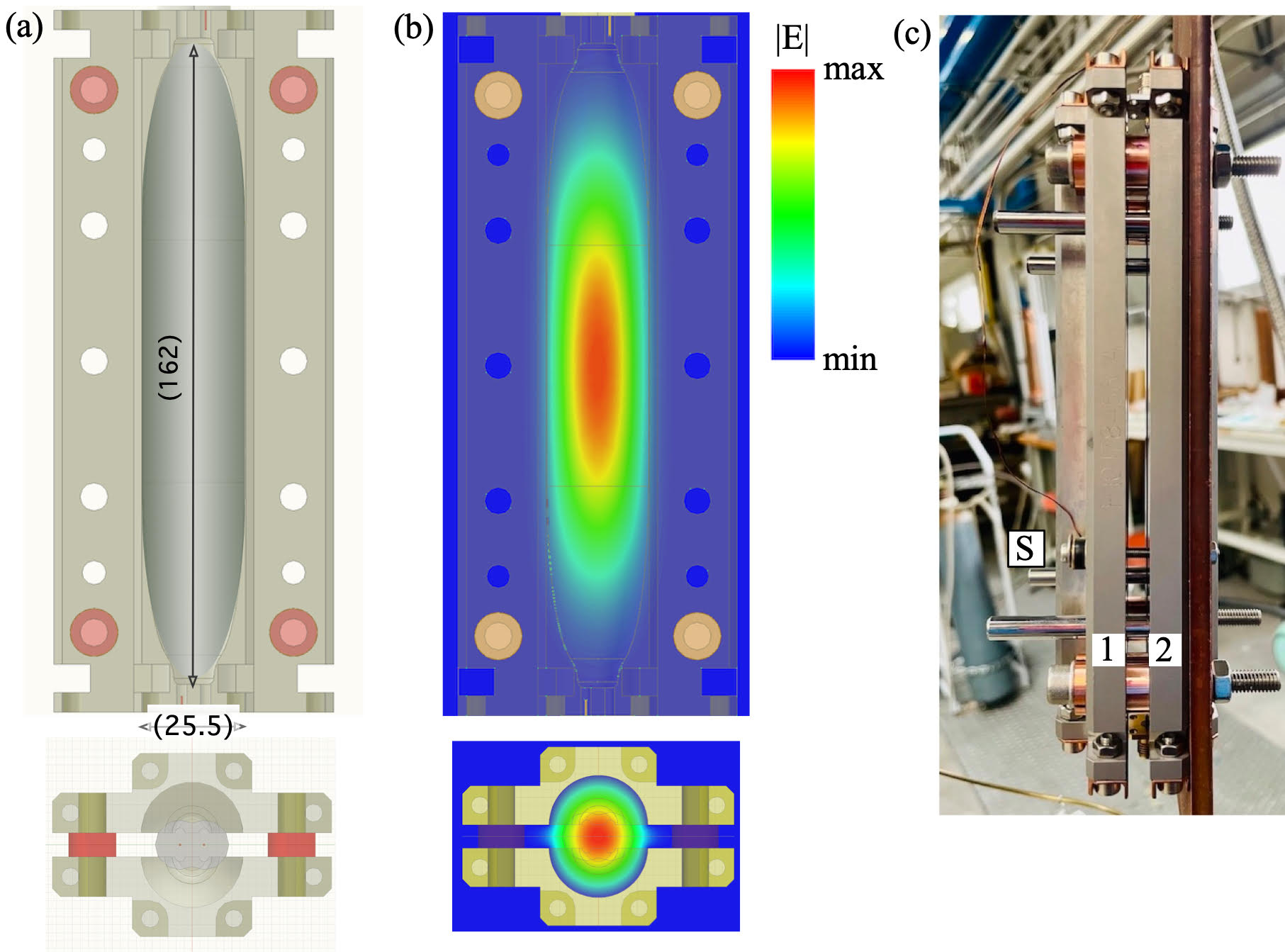}
    \caption{(a) Top view of one semi-cell and cavity cross section, with dimensions, in mm, along the relevant axes. The parts in red are copper spacers used to separate the two cavity halves. The lateral plates are colored in light gray, while the cigar-shaped inner surface is shown in darker gray. (b) TM$_{010}$ electric field mode profile obtained with FEM Simulations in the open configuration (6\,mm-gap). (c) Side view of the open cavity with 6\,mm gap. Sensor S measures the temperature of cavity half 1 thermalized to cavity half 2 through 4 copper spacers. Cavity half 2 is in direct contact with the copper cold finger.}
    \label{fig:1}
\end{figure}

Building on the demonstration of a Nb$_3$Sn cigar-shaped coated cavity for axion searches that resonates near $4\,\mathrm{GHz}$~\cite{Posen:2023}, we tested a rescaled cavity at $f_\mathrm{c}=9\,\mathrm{GHz}$, shown in Fig.~\ref{fig:1}. The fundamental TM$_{010}$ mode is selected as the science mode. The cavity is fabricated by carving the volume from two solid blocks of niobium; the two halves feature lateral plates on either side of the carved cavity region. The internal surfaces of both the cavity and the plates are electropolished and coated with Nb$_3$Sn using a vapor-diffusion process~\cite{Posen:2023}. We conduct experimental tests and finite-element method (FEM) simulations to investigate a tuning method recently introduced in the context of haloscope searches. We further develop the tuning-by-opening technique and extend its application to superconducting cavities, demonstrating a tuning range exceeding $1\,\mathrm{GHz}$ while maintaining $Q_0 \gtrsim Q_{\mathrm{DM}}$ across the entire explored frequency range, where $Q_{\mathrm{DM}}\sim 10^6$ is quality factor of the wave-like DM \cite{Turner:1990}. The lateral plates play a key role in maintaining a high Q even for large openings owing to the confinement effect that they provide for the electromagnetic axion sensitive mode.

\section{\label{sec:resistance} Surface resistance study}

\begin{figure}[ht]
    \centering
    \includegraphics[width=\linewidth]{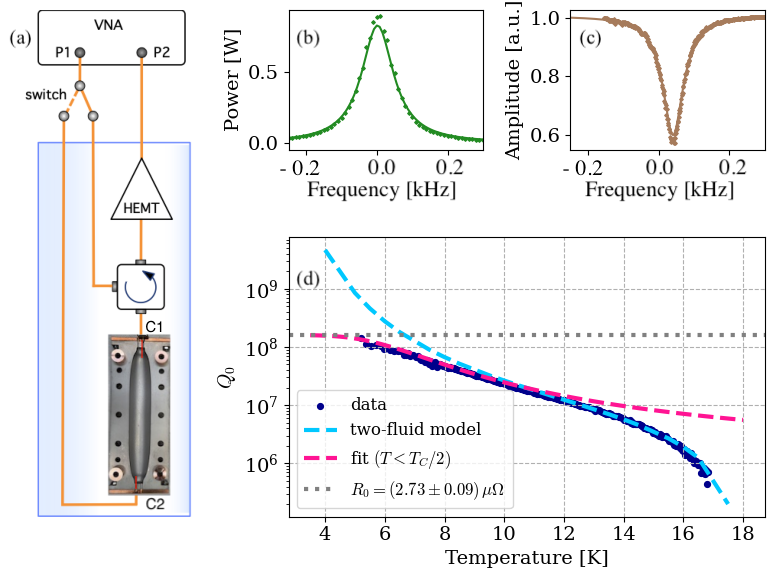}
    \caption{(a) Experimental setup. The cryogenic parts at approximately $4\,$K are the cavity, a circulator and a HEMT amplifier. A room-temperature switch routs the VNA output either to the weakly coupled antenna C2 for transmission measurement $S_{21}$ (b) or to the circulator port for measuring the reflection $S_{22}$ at the coupler C1 (c). In (b) and (c) data are fitted using Eqs.\, \ref{eq:1} and \ref{eq:2}, respectively. In both plots the resonant frequency $f_{\mathrm{c}} = 8.996217\,$GHz has been subtracted from the x‑axis. (d) Unloaded quality factor $Q_0$ measured at different temperatures below $T_C$, extracted by fitting the magnitude of the reflection function. The pink curve given by Eq.\,\ref{eq:7} fits the data for $T < T_C/2$. At low temperatures, the residual resistance $R_0$  dominates over the BCS resistance $R_{\mathrm{BCS}}$, which is exponentially suppressed. The reported $R_0$ is obtained from the fit. The light blue curve is obtained from surface resistance calculations based on BCS theory. The two-fluid model for the superconducting state \cite{Halbritter:1974} is used with the following parameters: energy gap $\Delta/k_\mathrm{B} \,T_\mathrm{C} = 2.06$, critical temperature $T_\mathrm{C} = 17.8\,$K, London penetration depth $\lambda_\mathrm{L} = 40\,$nm, coherence length $\xi_0 = 20\,$nm, and residual resistance ratio $\mathrm{RRR} = 1.74$.}
    \label{fig:2}
\end{figure}

For testing, the cavity is cooled via thermal contact with the cold finger of a cryocooler system at 3.6\,K. The cavity is equipped with two undercoupled dipole antennas: C1 is mounted on the top side, with coupling coefficient $\beta \sim 0.3$, while C2 is located on the bottom side and is more weakly coupled, with a coupling coefficient of $\sim 0.01$. We use a vector network analyzer (VNA) to measure the transmission $S_{21}$ and reflection $S_{22}$ parameters of the fundamental TM$_{010}$ mode, as shown in Fig.\,\ref{fig:2}\,(a).
For the $S_{21}$ measurement, the VNA sends a signal to C2, the transmitted signal passes through a circulator and is then amplified by a high-electron-mobility transistor (HEMT) amplifier before being fed into port P2 of the VNA.
For the $S_{22}$ measurement, an external switch redirects the VNA output from P1 to C1 through the circulator ports, while the reflected power at C1 is routed to P2.

The loaded quality factor $Q_\mathrm{L}$ is extracted from the transmission measurement by Lorentzian fitting $S_{21}$:
\begin{equation}\label{eq:1}
   S_{21}(f) = N_1\left|\frac{f_{\mathrm{c}}}{{f_{\mathrm{c}}}-2\,Q_{\mathrm{L}}\,i(f-f_\mathrm{c})}+i\,\alpha_1\right|+k;
\end{equation}
where $f_\mathrm{c}$ is the resonant frequency of the cavity mode, $N_1$ is a normalization factor, $\alpha_1$ accounts for possible impedance mismatches, and $k$ compensates for any offset.
The cavity frequency $f_\mathrm{c}$, the coupling coefficient $\beta$ of the coupler C1 and the unloaded quality factor $Q_0$ are obtained from the reflection parameter $S_{22}$:  
\begin{equation}\label{eq:2}
    S_{22}(f) = N_2\left| \frac{\beta - 1 + i Q_0 \left( \frac{f}{f_\mathrm{c}} - \frac{f_\mathrm{c}}{f} \right)}{\beta + 1 + i Q_0 \left( \frac{f}{f_\mathrm{c}} - \frac{f_\mathrm{c}}{f} \right)} +i\,\alpha_2\right|;
\end{equation}  
where $N_2$ and $\alpha_2$ are the normalization and impedance mismatch fitting parameters, respectively. The relation between the unloaded and loaded quality factors is $Q_0 = (1 + \beta)\,Q_{\mathrm{L}}$. 

To assess the present Nb$_3$Sn film quality, we estimate the surface resistance from quality factor measurements performed with the closed cavity at various temperatures. The intrinsic quality factor $Q_0$ relates to the surface resistance $R_\mathrm{S}$ as:  
\begin{equation}\label{eq:3}
    Q_0 = \frac{G}{R_\mathrm{S}},
\end{equation}  
where $G$ is a geometrical factor \cite{Padamsee} computed through simulations.

As there is no active temperature control system, we turn off the cryocooler and monitor the temperature with a sensor in thermal contact with the cavity as shown in Fig.\ref{fig:1}. The temperature initially changes as the square root of time and, after approximately $200\,$s, increases linearly at a rate of $0.38\,$K/min. During each measurement the temperature can thus be considered constant and homogeneous, since the VNA sweep time is set to $5\,$s. A total of $240$ reflection measurements are acquired in the temperature range from $5.2\,$K to $16.9\,$K. An accurate estimation of $Q_0$ at lower temperatures is not possible due to mechanical vibrations related to the pulse tube, that persist for about $30\,$s after shutdown.
Fig.\,\ref{fig:2}\,(d) presents the extracted values of $Q_0$ as a function of temperature. 
In the low-temperature regime, for $T < T_\mathrm{C}/2$, where $T_\mathrm{C}$ is the critical temperature of the Nb$_3$Sn film, the total surface resistance follows:  
\begin{equation}\label{eq:4}
    R_\mathrm{S}(T) = R_0 + R_\mathrm{{BCS}}(T).
\end{equation}  
$R_0$ is the temperature-independent residual resistance, that dominates the surface resistance of the Nb$_3$Sn film when approaching 0\,K. The primary contributions to $R_0$ include impurity scattering, film defects, trapped magnetic flux, and moving flux lines. The BCS (Bardeen–Cooper–Schrieffer) surface resistance depends on the temperature as: 
\begin{equation}\label{eq:5}
    R_\mathrm{{BCS}}(T) = A_\mathrm{S}\,f_\mathrm{c}^2\, \mathrm{e}^{-\frac{\Delta(T)}{k_\mathrm{B}\,T}},
\end{equation}  
where $k_{\mathrm{B}}$ is the Boltzmann constant, $A_\mathrm{S}$ is a function of material parameters, such as the Fermi velocity, the London penetration depth, the coherence length and the mean free path of electrons \cite{Padamsee}. $A_\mathrm{S}$ can be considered temperature-independent for $T < T_\mathrm{C}/2$. The function $\Delta(T)$ represents the energy gap and, for $T < T_\mathrm{C}/2$, it holds that $\Delta(T)\simeq\Delta(T=0)$, hence the energy gap in the following discussion is treated as a constant, $\Delta$.

Using Eq.\,\ref{eq:3}, Eq.\,\ref{eq:4} and Eq.\,\ref{eq:5} the measured $Q_0$ as a function of $T$ in the regime $T < T_\mathrm{C}/2$ is fitted with:  
\begin{equation}\label{eq:7}
    Q_0 = \frac{G}{R_0 + A\,\mathrm{exp}\big(-\frac{\Delta}{k_{\mathrm{B}}\,T}\big)},
\end{equation}  
where $G = 437\,\Omega$ is determined from FEM calculations, while $R_0$, the normalization factor $A$, and $\Delta$ are fitting parameters. The best-fit results yield $R_0 = (2.73 \pm 0.09)\,\mu \Omega$ and $\Delta = (5.0\pm 0.3)\,\times 10^{-22}$\,J. Assuming $T_\mathrm{C} = 17.8\,$K, corresponding to the lowest measured temperature before the superconducting-to-normal transition, we obtain $\Delta/(k_{\mathrm{B}}\,T) = (2.1 \pm 0.1)$, in agreement with typical values reported in the literature \cite{Padamsee}.
\section{\label{sec:FEM} Finite-element analysis}
\begin{figure}[t]
    \centering
    \includegraphics[width=\linewidth]{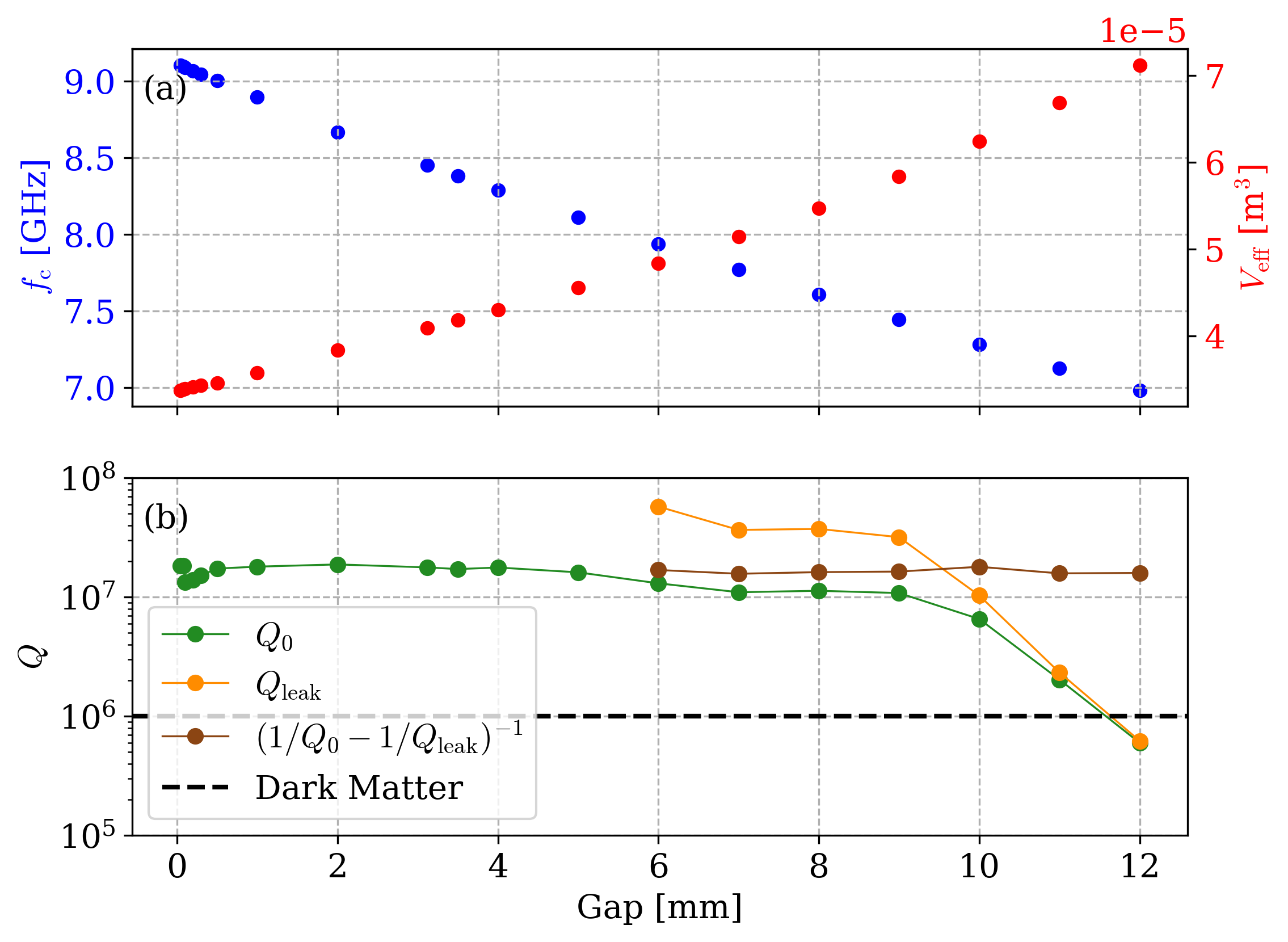}
    \caption{Simulation results. (a) TM$_{010}$ mode resonant frequency $f_\mathrm{c}$ (blue dots) and effective cavity volume (red dots) as function of gap size. (b) Unloaded quality factor $Q_0$ for different gap widths (green/orange dots), obtained by assigning finite/perfect conductivity to both the inner surface and lateral plates of the cavity. The chosen finite value of conductivity is $10^3$ times the copper conductivity at LHe temperature ($\sigma_{\mathrm{Cu}}^{4\mathrm{K}} = 3.2 \times 10^8\,$S/m). Computed (brown dots) intrinsic quality factor $(1/Q_0 - 1/Q_{\mathrm{leak}})^{-1}$ .}
    \label{fig:3}
\end{figure}
The electromagnetic behavior of the cavity and its tuning-by-opening is simulated in 3D using the finite-element method (FEM), in which the geometry is discretized into small elements forming a mesh, and Maxwell’s equations are solved for each element. The simulations are conducted with the Ansys HFSS \cite{hfss} suite in driven modal mode, which computes both the frequency response of the structure and the stationary field distribution when the system is excited by monochromatic signals injected through a port.  
In our model, the two antennas used experimentally for the transmission and reflection measurements are included explicitly, allowing us to obtain the corresponding $S$-parameters.
We simulate the full cigar-shaped geometry of the cavity, and we study the evolution of the electromagnetic field distribution as the two halves of the cavity are mechanically separated step by step using four 12\,mm - diameter copper ring spacers. The separation between the two half is changed by increasing the thickness of the four spacers. The cavity model used in the simulations is shown in Fig.\,\ref{fig:1}.  
Since the cavity is physically opened to perform the tuning, the electromagnetic field is not fully confined to the interior volume. Therefore, in the simulation, the cavity is enclosed within an external rectangular vacuum region. On the boundary of this outer region we apply radiation boundary conditions, which approximate an open space by absorbing normal or near-normal incident waves.
We assign to the inner cavity surface and to the lateral plates an effective conductivity about $10^3$ times the copper conductivity at LHe temperature\cite{Calatroni2020} ($\sigma_{\mathrm{Cu}}^{4\mathrm{K}} = 3.2 \times 10^8\,$S/m), that produces an unloaded quality factor of approximately $2\times10^{7}$. The four spacers are included in the simulation as well, with conductivity $\sigma_{\mathrm{Cu}}^{4\mathrm{K}}$.

The cavity response excited through the antenna is studied in the frequency region where the axion-sensitive $\mathrm{TM}_{010}$ mode lies. The effective cavity volume $V_{\mathrm{eff}} = C_{010}\,V$ is computed by the simulation, where $C_{010} \approx 0.5$ is the cavity form factor\cite{Sikivie2021}.
The simulated cavity frequency $f_\mathrm{c}$, $V_{\mathrm{eff}}$ and the unloaded quality factor $Q_0$ for different gap values are shown in Fig.\,\ref{fig:3}. 

For openings below $0.5\,\mathrm{mm}$, additional modes arise that are confined in the region between the two facing lateral plates, and they may mix with the axion-sensitive mode, leading to a degradation of $Q_0$ by about $30\%$.
To capture this behavior, the mesh is locally refined on the lateral plates down to sub-$0.5\,\mathrm{mm}$ resolution.
For thicker spacers ($0.5$ and $1\,\mathrm{mm}$), the measured $Q_0$ stabilizes around $2\times10^{7}$.  
For this reason, the same unloaded quality factor was imposed on the FEM model by assigning a suitable effective conductivity to the coated cavity surfaces.
The $Q_0$ value remains nearly constant up to approximately $5\,\mathrm{mm}$ plate separation, beyond this point it begins to decrease, losing roughly a factor of two between $5$ and $9\,\mathrm{mm}$, and then degrading by more than one order of magnitude between $9$ and $12\,\mathrm{mm}$.
This latter behavior is attributed to direct radiative losses set by $Q_\mathrm{r}$, related to power leaking from the cavity through the increasing aperture into the surrounding simulation volume.
A phenomenological model that can account for different contributions to the measured quality factor can be written as:
\begin{equation}
\frac{1}{Q_0} 
= \frac{1}{Q_{\mathrm{int}}}
+ \frac{1}{Q_{\mathrm{lp}}}
+ \frac{1}{Q_{\mathrm{leak}}},
\end{equation}
where $Q_{\mathrm{int}}$ is the intrinsic quality factor of the cavity when closed, set by the surface resistance of the Nb$_3$Sn film of the inner surface. We introduce $Q_{\mathrm{lp}}$ to account for losses due to the poor film quality on the lateral plates. $Q_{\mathrm{lp}}$ is expected to vary with the gap, since an increasing aperture exposes a progressively larger area of the plates to the RF fields. Finally the term $Q_{\mathrm{leak}} = \left( \frac{1}{Q_\mathrm{r}} + \frac{1}{Q_{\mathrm{mm}}} \right)^{-1}$, describes losses due to power leakage from the cavity when it is opened, with $Q_{\mathrm{mm}}$ associated with indirect radiative losses through mixing with modes confined between the lateral plates.

One of the datasets presented in Fig.\,\ref{fig:3} reports results obtained by assigning perfect conductivity to both the inner cavity walls and the lateral plates. These points therefore represent the contribution of $Q_{\mathrm{leak}}$ to the total quality factor.
We observe that the trend of $Q_0$ follows that of $Q_{\mathrm{leak}}$ for apertures larger than approximately $9\,\mathrm{mm}$, where reasonably $Q_{\mathrm{leak}} \simeq Q_\mathrm{r}$, indicating that direct radiative leakage through the opening is the dominant mechanism limiting the quality factor in that region.

As for the required mechanical tolerances in this system, we perform simulations taking into account misalignments between the two cavity halves.
\begin{figure}[t]
    \centering
    \includegraphics[width=\linewidth]{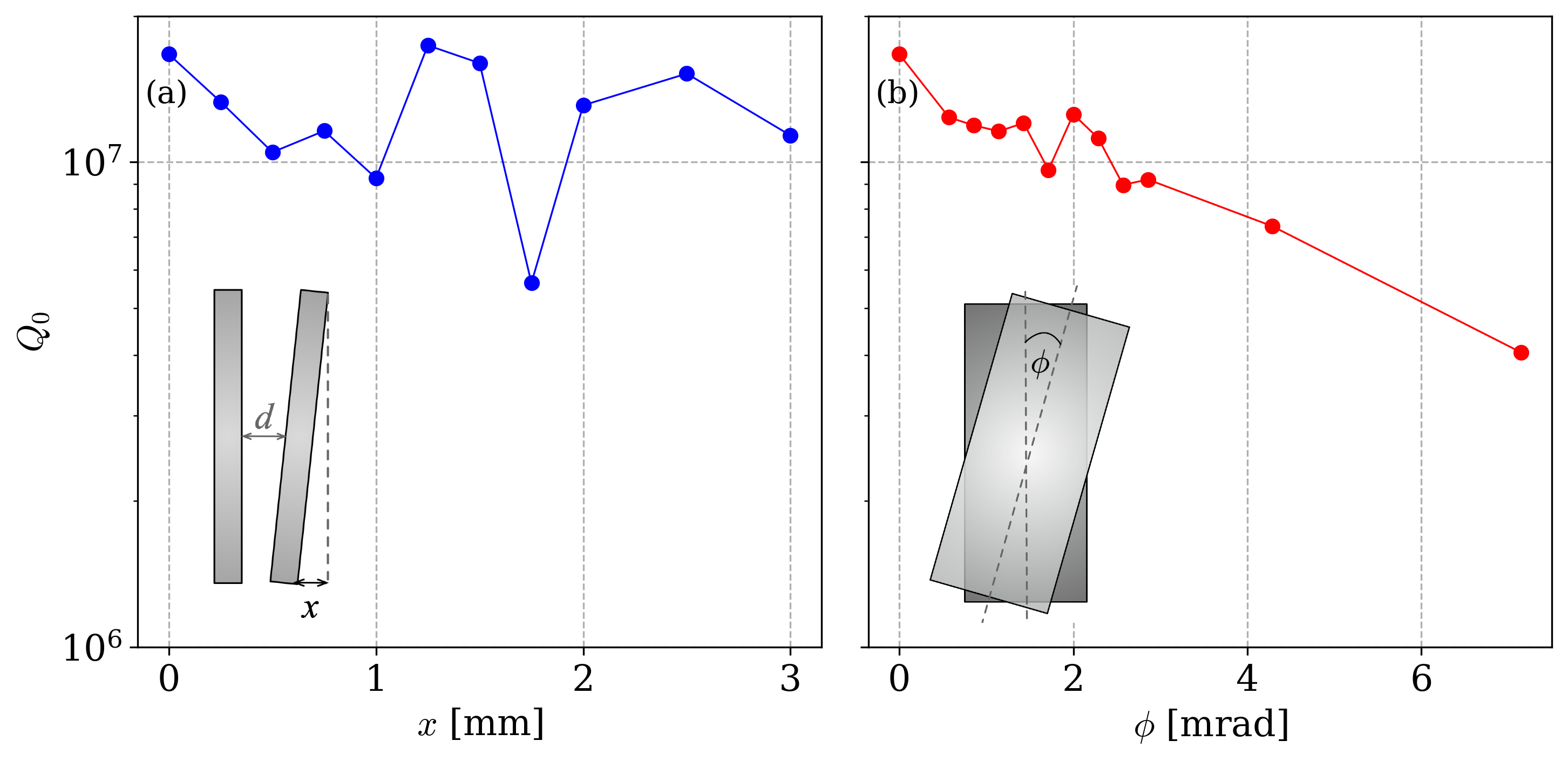}
    \caption{Impact of misalignments between the two cavity halves on $Q_0$. In the FEM simulations we consider both (a) the possible lateral misalignment $x$ relative to the parallel configuration and (b) the effect of a rotation angle $\phi$, as depicted in the insets.
    }
    \label{fig:4}
\end{figure}
In the simulations we introduce the misalignment parameter $x$, corresponding to a deviation from the parallel configuration in which the short side of one cavity half is displaced toward the other, and the rotation angle $\phi$ between the two halves, as illustrated in Fig.~\ref{fig:4}. We calculate the unloaded quality factor at gap width $d = 2\,$mm for different values of $x$ and $\phi$.
$Q_0$ gradually decreases as a function of the rotation angle, whereas no clear dependence on $x$ is observed. Notably, the decrease of the effective volume $V_{\mathrm{eff}}$ due to the $x$ misalignment is below 3\%, while the rotation angle $\phi$ do not affect it at all.  
 
In summary, simulations show that with this tuning-by-opening method it is possible to tune the cavity without significantly impacting on $Q_0$, up to a gap width comparable to the cavity radius, corresponding to a 20\% tuning at 9\,GHz. 
The required mechanical tolerances are not too stringent and look experimentally achievable.

\section{Experimental Characterization of the Tuning Mechanism}

\begin{figure}[t]
    \centering
    \includegraphics[width=\linewidth]{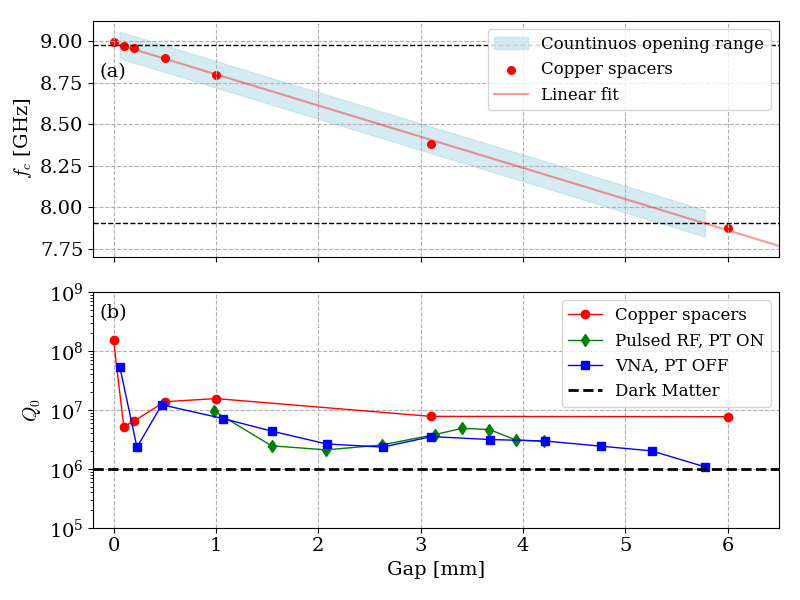}
    \caption{(a) Tuning slope of the resonant frequency obtained using copper spacers corresponding to several gap widths; the closed-cavity value is also included. The linear fit yields a tuning slope of $188\,\mathrm{MHz/mm}$. (b) Unloaded quality factor $Q_0$ extracted from reflection measurements performed with spacers (red circles). The green diamonds and blue squares correspond to measurements obtained with the continuous tuning system, with the green dataset derived from the cavity decay time constant measured using the pulsed RF method at selected frequencies and the blue dataset obtained with a VNA. The pulsed RF method enables characterization even when the pulse tube (PT) is operating. The errors assigned to each measurement is not visible since it is within the symbols
}
    \label{fig:5}
\end{figure}

To demonstrate the proposed tuning mechanism, we perform preliminary tests using copper spacers inserted between the two cavity halves. Their thicknesses range from 0.1 to 6.0\,mm corresponding to frequencies from $9.0\,$GHz down to $7.9\,$GHz, as shown in Fig.\,\ref{fig:5}. 
A drop of more than one order of magnitude is observed as soon as the cavity is opened, followed by a slightly steeper dependence of $Q_0$ on the gap width than predicted by the simulations.
This could be ascribed to the low quality of the Nb$_3$Sn film on the lateral plates due to scratches and non-uniformity, an effect not included in the FEM model, which assumes identical surface properties for the inner cavity walls and the lateral plates. Despite this modeling assumption, the preliminary tests with spacers validate the simulated trends, and even for the largest tested gap of $6\,$mm we still measure $Q_0 \approx 8\times 10^6$. Therefore a mechanical sliding system to enable continuous frequency adjustment under cryogenic conditions is implemented.

In this set up, one cavity half is anchored to a copper L-shaped bracket in thermal contact with the $3\,$K plate, while the other half is mounted on a movable cart. The cart is attached to a second copper plate, also thermalized at $3\,$K. Translation of the movable cavity half is achieved through a fiberglass rod connecting the cart to a room‑temperature positioner, and its thermalization is ensured by copper braids. The temperatures of the two halves are monitored by two Cernox thermometers.

\begin{figure}[t]
    \centering
    \includegraphics[width=\linewidth]{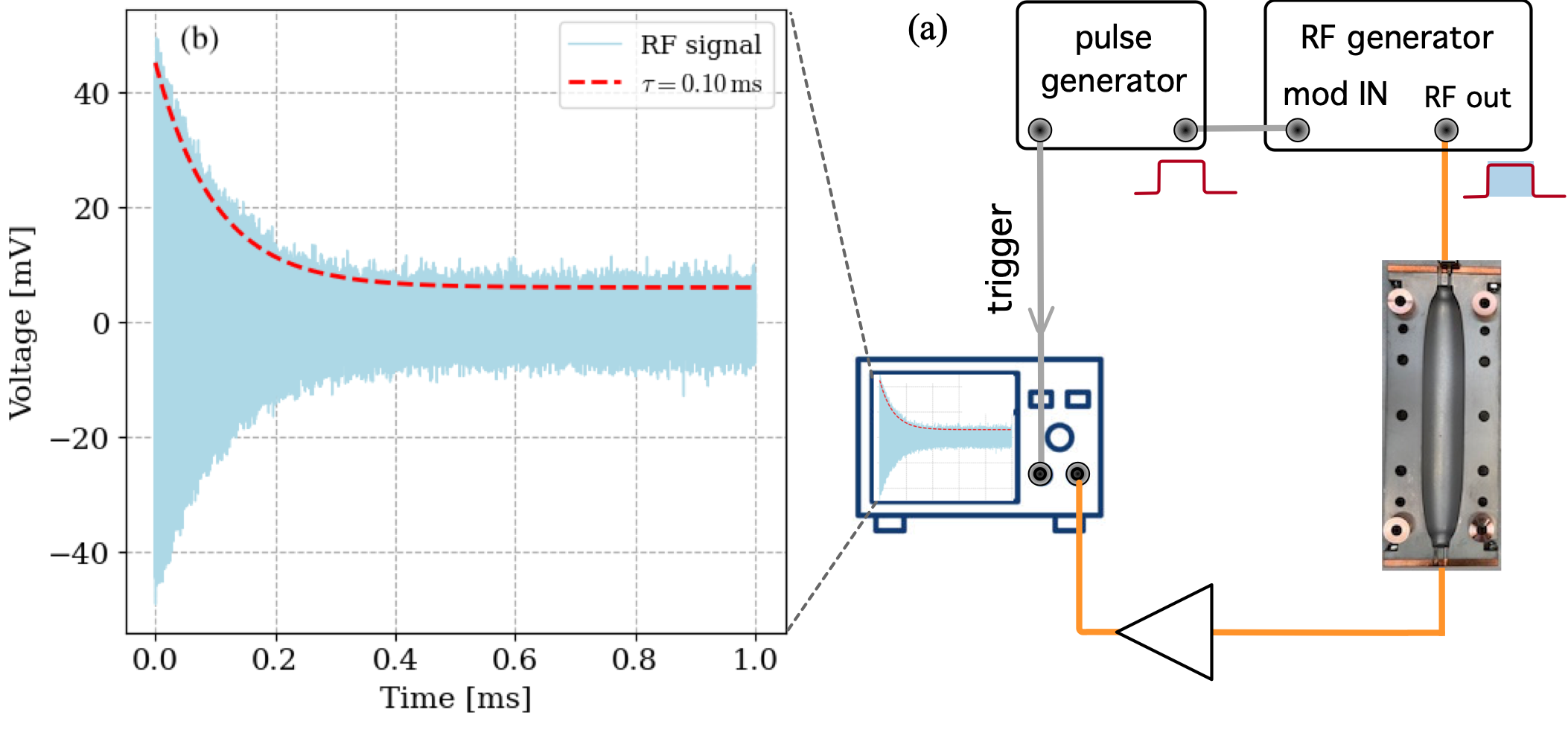}
    \caption{(a) A pulse generator provides the modulation input to a microwave generator in the form of a $800\,\mu\mathrm{s}$-duration square pulse. The pulsed RF signal with frequency $f_\mathrm{c}$ is sent to the cavity, whose transient response (b) is recorded with a 20\,GHz digital sampling oscilloscope. The dashed line is an exponential fit of the signal envelope.}
    \label{fig:6}
\end{figure}

The continuous tuning mechanism is tested in two configurations, corresponding to measurements performed in the time and frequency domain. Pulsed RF measurements are employed when vibrations originating from the soil and the pulse tube prevent a reliable determination of $Q_0$ with a VNA. 
As shown in Fig.\,\ref{fig:6} we drive the cavity with $800\,\mu\mathrm{s}$-duration microwave square pulses and record its response at a 20\,GHz digital sampling oscilloscope.

For eight chosen apertures we acquire several cavity transient decays that yield the cavity loaded quality factor $Q_\mathrm{L}=\pi f_\mathrm{c} \tau$, with $\tau$ obtained by fitting the exponential decay.
Compared to the frequency domain measurements using a VNA, this pulsed RF method allows for the determination of the quality factor in the presence of mechanical vibrations that couple to the system.
During these tests the tuning mechanism can be adjusted from 1.0 to 4.2\,mm gap, and the measured temperatures of the fixed and movable cavity halves are 4.5\,K and 5.5\,K, respectively.

We implement acoustic‑noise mitigation strategies, including vibration‑damping layers between the laboratory floor and the cryostat and Teflon supports that enclose the antennas to reduce vibration‑induced motion. Soil-borne vibrations are suppressed, and acoustic noise associated with the pulse tube is reduced. As a result, frequency-domain measurements using a VNA can now be performed when the pulse tube is switched off.

As we do not precisely control the gap width when using the sliding mechanism, we infer its value from the set frequency using the tuning slope of 188\,MHz/mm obtained from the tests with copper spacers.

In Fig.~\ref{fig:5} we compare the unloaded quality factors obtained at different frequencies with the two methods, where for the pulsed RF technique $Q_0$ is computed using the coupling factor $\beta$ measured with the VNA. Across the entire investigated frequency range we observe compatible results within the systematic uncertainties associated with mechanical tolerances of the tuning system, with $Q_0$ consistently exceeding the DM quality factor.

A reduction in  quality factor is observed in these datasets when compared with the simulation results. Moreover, for openings above 0.5\,mm the $Q_0$ measured with the sliding mechanism is smaller than the value obtained using the copper spacers tuning method. Modes arising in the region between the lateral plates are expected to have a stronger impact when the cavity halves are misaligned. Hence, this difference in $Q_0$ is possibly due to misalignment between the two cavity halves, which is expected to affect the continuous tuning configuration, as previously observed also for copper cavities with much lower quality factor \cite{Golm2024}. 

Additionally, the discrepancy between the simulations results and the data obtained when using the tuning-cart mechanism may also originate from the markedly different boundary conditions in the two cases: in the first, copper spacers are placed next to the cigar, whereas in the continuous tuning configuration the cart’s metallic plate lies only at a few millimeters from the cavity external edge.


\section{\label{sec:concl} Discussion and conclusions}

We have demonstrated a continuous tuning method applicable to high-$Q$ superconducting microwave cavities for haloscope DM searches. Finite-element simulations of a model based on an existing cigar-shaped Nb$_3$Sn cavity show that separation of the two cavity halves allows for continuous tuning from $9.0\,\mathrm{GHz}$ to $7.5\,\mathrm{GHz}$ without degrading the quality factor, which remains at the level of $10^7$ across the entire range. This tuning span corresponds to an $9\,\mathrm{mm}$ gap between the two halves. Above this value, radiative losses $Q_\mathrm{r}$ become significant, as demonstrated by a model with perfectly conducting boundaries, as shown in Fig.~\ref{fig:3}.

Preliminary experimental tests, conducted with copper ring spacers for cavity halves separation, motivated the development of an {\it in-situ} sliding mechanism, that provides continuous tuning over $1\,$GHz, with a quality factor that remains higher than $Q_{\mathrm{DM}} \sim 10^6$ throughout the full frequency range, as reported in Fig.\,\ref{fig:5}.
Mode crossings commonly affect rod-based tuning mechanisms, thereby limiting the usable fraction of the accessible frequency range. In contrast, with the tuning-by-opening approach, no mode crossings are observed over the entire tuning range, in agreement with expectations \cite{Braggio:2023}.

Moreover, it should be possible to obtain even higher Q values, close to those predicted by simulations, using a cavity built specifically for this purpose. The cavity employed in this work has in fact a low-quality film on the lateral plates as it was not originally designed for the presented tuning-by-opening method. 

When the sliding mechanism is operated to separate the two halves, no direct control over rotations or deviations from parallel alignment is possible. Nevertheless we consistently obtained unloaded quality factors of $10^6$ or higher. This demonstrates the robustness of the tuning concept against mechanical asymmetries and validates the misalignment‑tolerance study shown in  Fig.\,\ref{fig:4}.

In a dilution refrigerator system, with the cavity installed at the mixing chamber plate, a better decoupling from pulse tube vibrations is expected.
In addition, a motorized actuator can be straightforwardly integrated into the mechanical tuning system to enable automated control of the cavity semi-cell position, synchronized with data acquisition.
Finally, this tuning method, which maintains $Q_0$ above $Q_{\mathrm{DM}}$ over a wide frequency range, can be easily applied to REBCO cavities that can reach $Q_0\sim10^7$ in a multi-tesla magnetic field \cite{Ahn:2022}, directly enabling high-Q, broadband axion searches.
These developments will allow the demonstrated cavity concept to be directly integrated into next-generation haloscope experiments.

\section{\label{sec:acknow} Acknowledgments}

We acknowledge technical support from E. Berto, A. Benato and M. Zago (INFN Padova) for the design and realization of the cavity tuning system. The contribution of M. Tessaro and F. Calaon (INFN Padova) to the experiment electronics and cryogenics, respectively, is gratefully acknowledged. 
We also acknowledge M. Checchin, I. Gonin, and O. Pronitchev (Fermilab) for their contributions to the cavity design.

This work was supported by the U.S. Department of Energy, Office of Science, National Quantum Information Science Research Centers, Superconducting Quantum Materials and Systems Center (SQMS), under Contract No. 89243024CSC000002. Fermilab is operated by Fermi Forward Discovery Group, LLC under Contract No. 89243024CSC000002 with the U.S. Department of Energy, Office of Science, Office of High Energy Physics.

\nocite{*}
\bibliography{biblio}

\end{document}